\documentclass[11pt]{article}
\usepackage{blois,epsfig}

\def\Journal#1#2#3#4{{#1} {\bf #2}, #3 (#4)}

\def\PRL{\em Phys. Rev. Lett.}

\def\be{\begin{equation}}
\def\ee{\end{equation}}
\def\bea{\begin{eqnarray}}
\def\eea{\end{eqnarray}}

\def \met {\mbox{${\not}{E_T}$} }
\begin{document}
\vspace*{4cm}
\title{ELECTROWEAK AND TOP PHYSICS AT THE TEVATRON AND THE LHC}

\author{ G. CHIARELLI }

\address{Istituto Nazionale di Fisica Nucleare, Sezione di Pisa, L.go Bruno Pontecorvo, 3\\
I-56127 Pisa, Italy}

\maketitle\abstracts{
In the last decades electroweak processes were studied at hadron and lepton colliders.
By exploiting the large statistics and the c.o.m. energy available, hadron colliders 
played a significant role in performing precision measurements of standard model 
parameters and in observing rare processes. Besides, in the last decade of the XX 
century, the last fermion predicted -the top quark- was discovered at the Tevatron
collider. We are now at the start of a new hadron collider, the LHC, and in this 
paper, I will review recent results from the Tevatron and compare perspectives for 
experiment taking data at the two accelerators.}

\section{Introduction}
The Tevatron Collider, where protons and antiprotons collide at $\simeq 2 
$ TeV of c.o.m energy,
operated at Fermilab (in the outskirts of Chicago) since the mid-eighties of the last 
century. Since then it has played a key role in studying the electroweak 
processes, and is also the place where,
in 1994-1995 the first observation, and then the discovery of the top 
quark, took place~\cite{topd}.
After a long shutdown, the machine and the two detectors (CDF and D0) were upgraded
to run at increased luminosities. Since 2001 CDF and D0 have been steadily 
taking data. 
Here I will present results obtained with an integrated luminosity 
of 2$\div$4 fb$^{-1}$~\footnote{1 fb$=10^{-39}cm^2$}.

The Large Hadron Collider (LHC) was built at CERN, and will collide 
protons on protons, 
starting in fall 2009. In the first year of operation, collisions 
will take place
at a c.o.m. energy of 7 TeV. After a shutdown it will increase the beam energy
to reach its design goal of 14 TeV.
Unlike the Tevatron, aimed to generically study processes predicted in the standard
model of elementary particles (SM) framework, the LHC goal is to 
specifically explore the electroweak symmetry breaking
process and to look for new physics. In order to fulfill this task, two full purpose
detectors (ATLAS and CMS) were built, along with two dedicated ones (LHCB, to 
flavour physics and TOTEM to diffraction and elastic processes). Finally, as the LHC will
also collide ions to study high-density states of matter, a fifth detector -ALICE- was built 
to this scope.
In the following, when discussing LHC perspectives, I will limit myself to 
future measurements by ATLAS and CMS.

\section{The Environment}
Both Tevatron and LHC share the common ground of colliding hadrons (proton and antiprotons
at the Tevatron, protons at the LHC). This implies that in the hard scattering between 
partons only the fraction of energy carried by the interacting quarks and gluons, is
available to produce interesting events. Most of the energy is
lost into peripheral ("soft") interactions. Moreover the soft interactions between the
hadrons is also responsible for most of the cross section. At the Tevatron, the total
inelastic cross section is about 50 mb, while, for example, single-top production
has a cross section of $\sim 3$ pb, or $\sim $3$\times$ 10$^{-9}$ mb (see 
figure~\ref{fig:cross} for a comparison of several production cross section). It 
is obvious that, in order to 
study processes at the pb rate, a trigger system, capable of rejecting more than 99.999\% of
the events, without introducing dead-time, is needed.
The LHC, with its even higher energy and instantaneous luminosity, provides similar challenges
to ATLAS and CMS in order to fulfill their physics goals.

\subsection{The Tevatron}
The Tevatron collider started operation in October 1985, recording an handful of events at CDF
(back then the only operational detector) at $\sqrt{s}=1.6$ TeV. After a first data taking 
period ("Run 0") in 1988-1989 at 1.8 TeV, the two detectors collected
$\simeq 120$ pb$^{-1}$ of data during the long (1992-1996) Run I. In this period a number of
striking measurements and discoveries were made. Most and foremost the top quark was first
observed (1994) and then discovered in 1995~\cite{topd}, but I would also 
like to mention the precision
measurements of the $W$ mass and many QCD studies and B-physics results.

\begin{figure}
\begin{minipage}{0.45\linewidth}
\psfig{figure=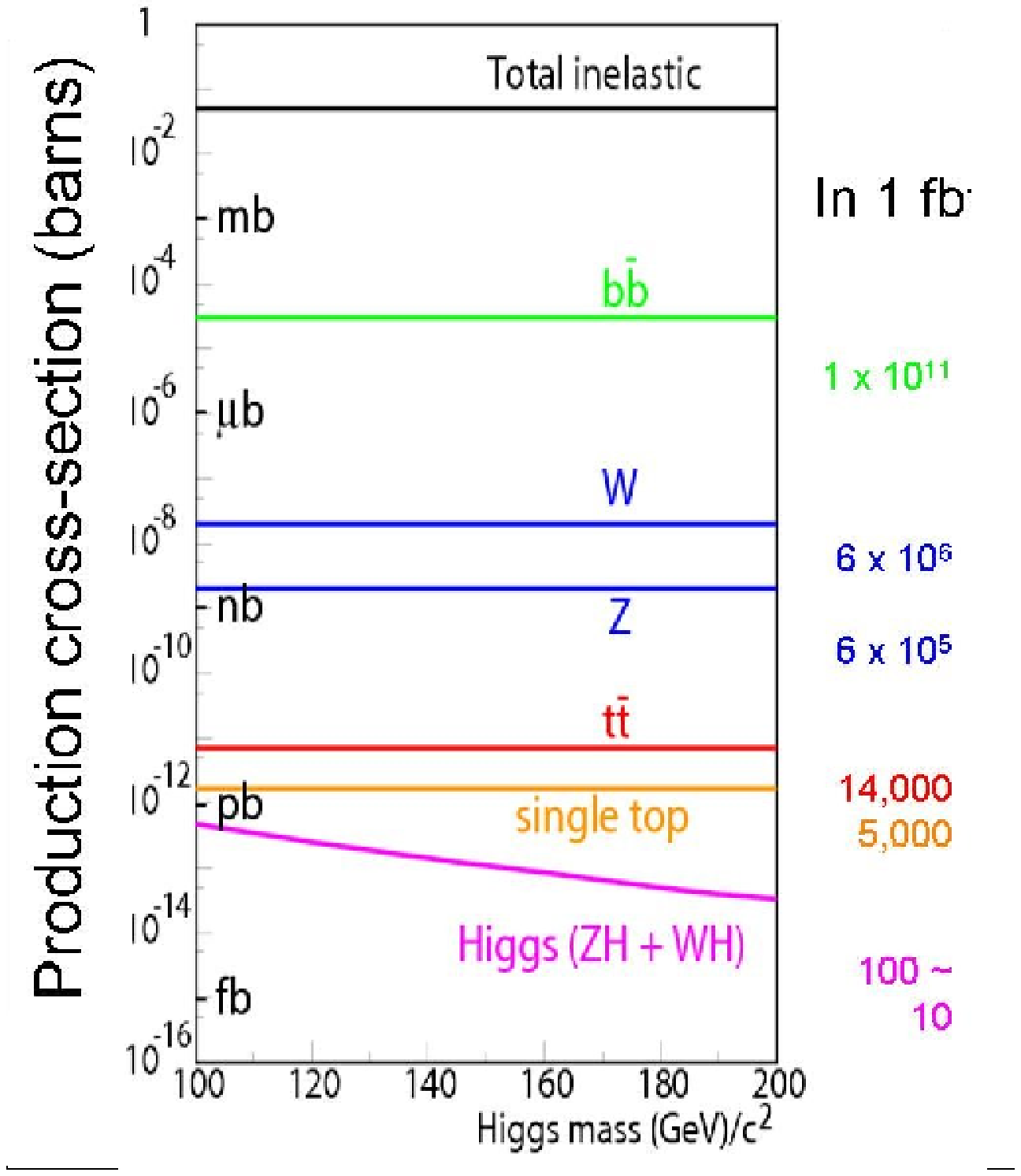,height=1.7in}
\caption{Cross section for relevant processes at the Tevatron. Cross section
involving Higgs plotted as a function of the Higgs mass.}
\label{fig:cross}
\end{minipage}
\begin{minipage}{0.45\linewidth}
\psfig{figure=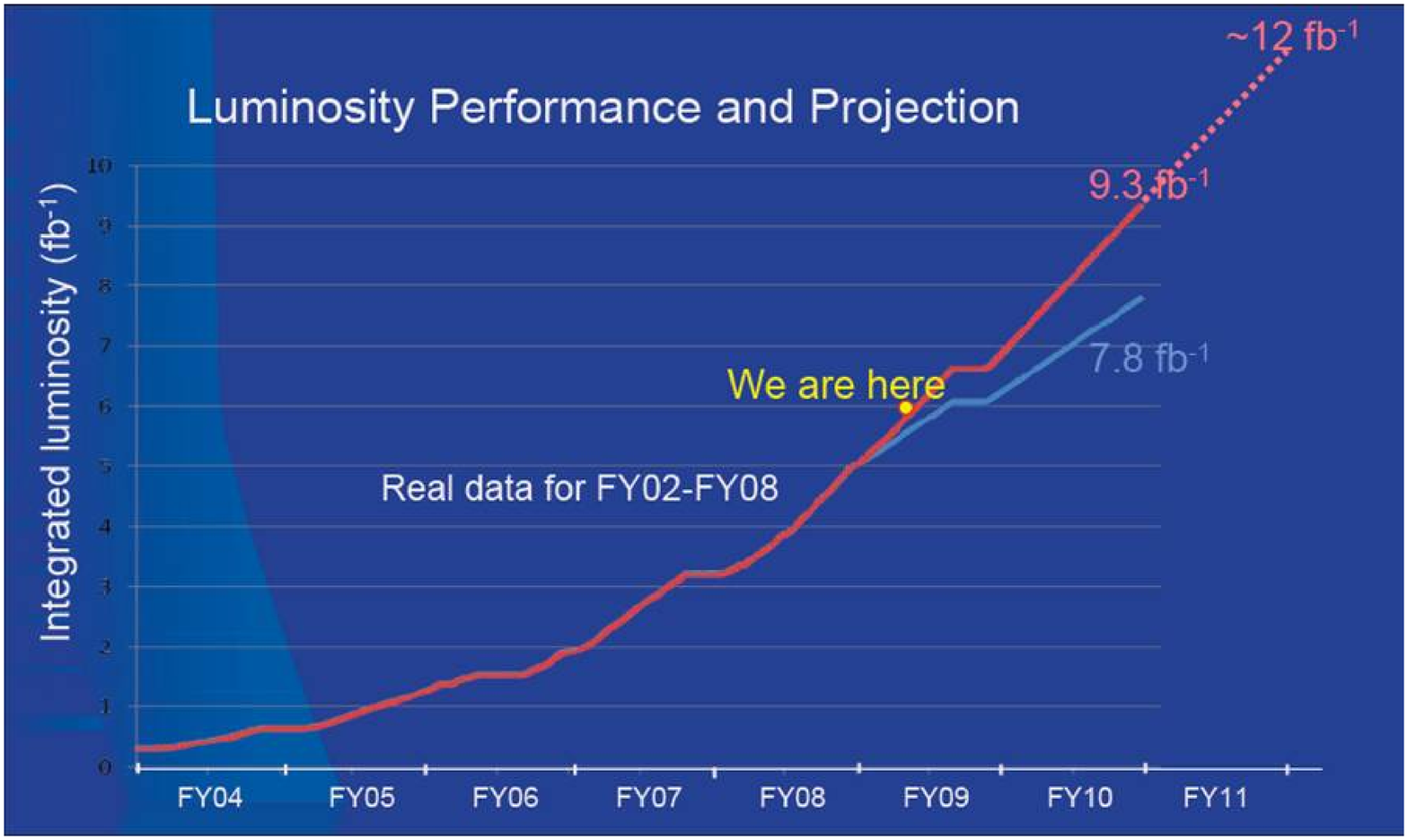,height=1.7in}
\caption{Run II integrated luminosity as a function of time. Past and future.}
\label{fig:lump}
\end{minipage}
\end{figure}

It was clear that, while originally designed to provide a maximum luminosity of 10$^{30}$ 
cm$^{-2}$ s$^{-1}$, an upgrade of the luminosity could extend the physics reach of
D0 and CDF and provide the chance to explore in deep the structure of the 
electroweak processes
as well as the nature of the ElectroWeak Symmetry Breaking (EWSB).
The improvement of the machine and, in parallel, the upgrades of the two 
detectors, brought to the start of Run II (2001) with new expectations.
Soon became clear that, instead of the 2 fb$^{-1}$ per experiment, 
as originally foreseen, more data
could be available by the time the LHC went into operation.
We are now in the eighth year of Run II and, after a shaky startup, the 
Tevatron
is now routinely colliding proton and antiproton at $\sqrt{s}$=1.96 TeV and 
instantaneous luminosity in excess of 10$^{32}$ cm$^{-2}$ s$^{-1}$.
These outstanding performances already delivered to each detector about 5 fb$^{-1}$
of data or (to put things in perspectives) more than 33K $t\bar{t}$ pairs, 30millions
$W$, more than 20,000 $WW$ etc. 
The current plan of the Laboratory foresees running until the end of FY 2011, in 
figure~\ref{fig:lump} I show the past history and the current prediction for the 
integrated luminosity.

\subsection{The LHC}
The LHC, built at CERN in the LEP tunnel, is a 23 Km diameter machine designed to collide
protons on protons at $\sqrt{s}$= 14 TeV at an instantaneous luminosity of
10$^{34}$ cm$^{-2}$ s$^{-1}$. The increase in c.o.m. energy, together with the very short
interbunch (25 ns) set an harsh environment for the detectors: each bunch crossing at
design luminosity is expected to see the overlap of (on the average)
$\sim $ 20 soft interactions. ATLAS
and CMS are designed to, first of all, detect the needle in the haystack 
and write to tape only a very tiny fraction of events where the hard 
scattering took place.
The first data taking run is expected to start in Fall 2009, at a c.o.m. energy of
7 TeV (each beam at 3.5 TeV) and last until about 100 pb$^{-1}$ have been 
recorded by each experiment~\cite{lhcs}.
 
\section{Electroweak Physics}
The cornerstones of electroweak physics are $Z$ and $W$ bosons. Copiously produced at the
Tevatron, they are routinely used as monitoring tools, both during data taking and
in testing the offline algorithms. Their leptonic decays in $e$, $\mu$ and $\nu$ 
provide easy trigger signatures.
Moreover, the very large number of events, allow the precision study of their properties.

\subsection{W and Z production}
\begin{figure}
\begin{minipage}{0.45\linewidth}
\psfig{figure=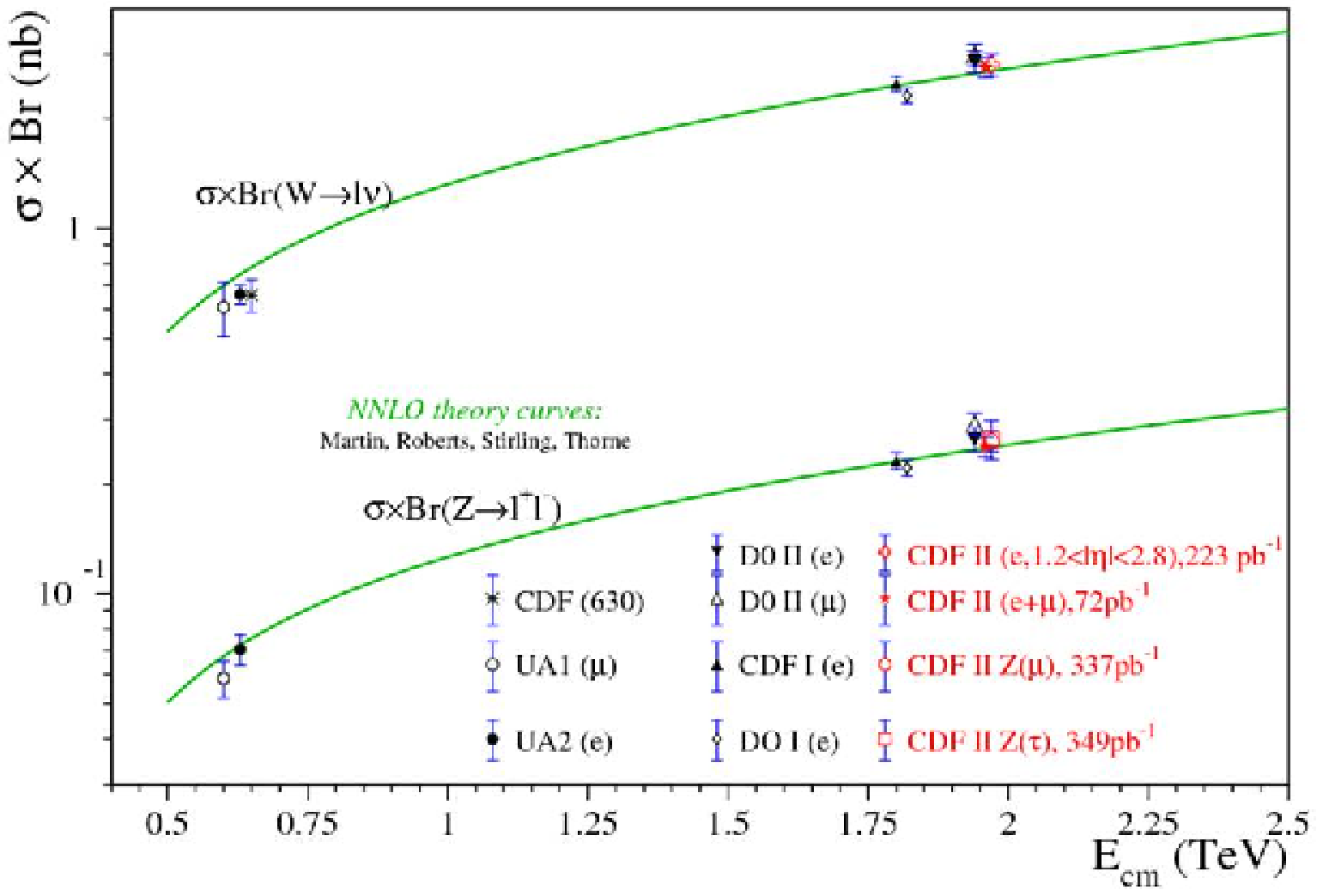,height=1.7in}
\caption{$W$ and $Z$ production cross section measured at hadron 
colliders.}
\label{fig:wzhad}
\end{minipage}
\begin{minipage}{0.45\linewidth}
\psfig{figure=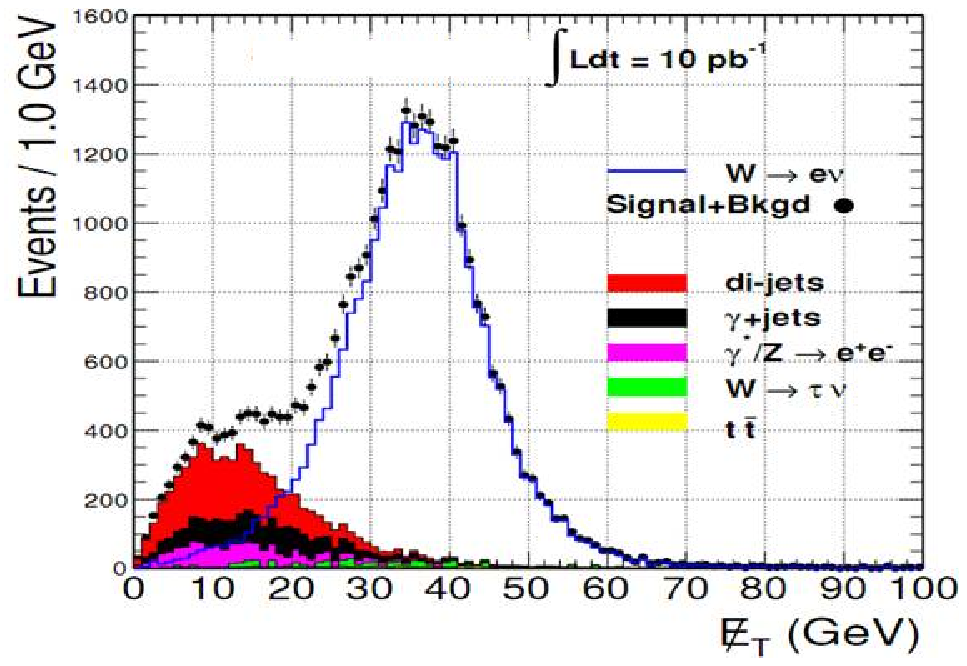,height=1.7in}
\caption{Missing Et distribution in $W$ candidates event at CMS.}
\label{fig:cmsw}
\end{minipage}
\end{figure}

An interesting milestone is the measurement of the cross sections and the 
comparison of data and theoretical prediction.
In figure~\ref{fig:wzhad} I show the compilation of the measurements at 
hadron colliders.
It will be interesting to compare the NNLO prediction with the LHC 
results as large increases are predicted as a function of $\sqrt{s}$. 
$W$ cross section is predicted to be 20(12) nb at 14 (10) TeV, while $Z$
production is 2 (1.2) nb at 14(10) TeV.
Figure~\ref{fig:cmsw} shows the expected \met distribution at CMS in 10 pb$^{-1}$ 
at 10 TeV. 
Expectation, after all cuts, is of the order of 10,000 events per 
experiment in 50 pb$^{-1}$ at 10 TeV.
ATLAS, with as few as 50 pb$^{-1}$ at 14 TeV, predicts $\Delta \sigma 
/\sigma \simeq$ 5 (3) \% in the $e$ ($\mu$) channel~\cite{atdr}.  

As $W$ and $Z$ cross sections brings information about the parton 
distribution functions (p.d.f.) of 
quarks and gluons, the $W$ asymmetry and the d$\sigma$/dy for $Z$ events provide 
constraint for the p.d.f. Those measurements, to be useful, need a good 
understanding of the backgrounds and of systematic effect. In a first 
phase, at LHC, might be more useful to perform a measurement of the ratio 
of forward-to central cross sections that can be directly compared to NNLO 
calculation and provide early information on the p.d.f.~\cite{forward}.

\subsection{Diboson Production}
\begin{figure}
\begin{minipage}{0.45\linewidth}
\psfig{figure=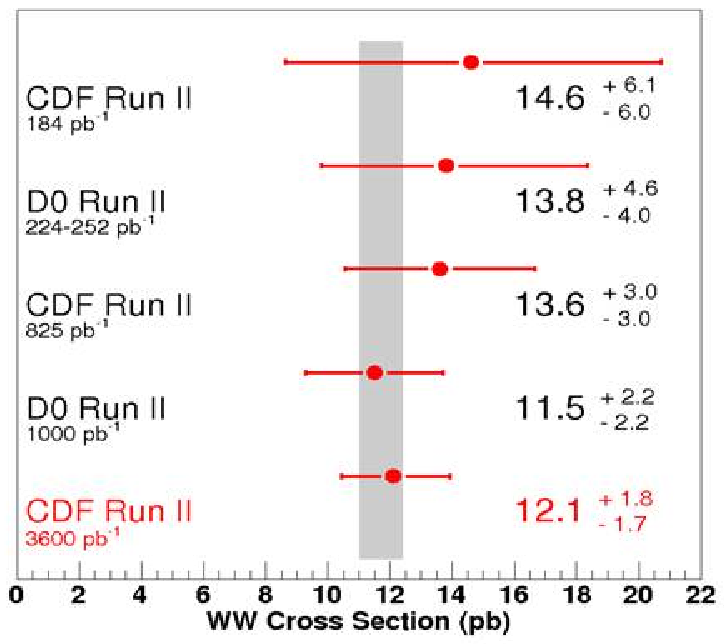,height=1.7in}
\caption{$WW$ production cross section at CDF.}
\label{fig:dibt}
\end{minipage}
\begin{minipage}{0.45\linewidth}
\psfig{figure=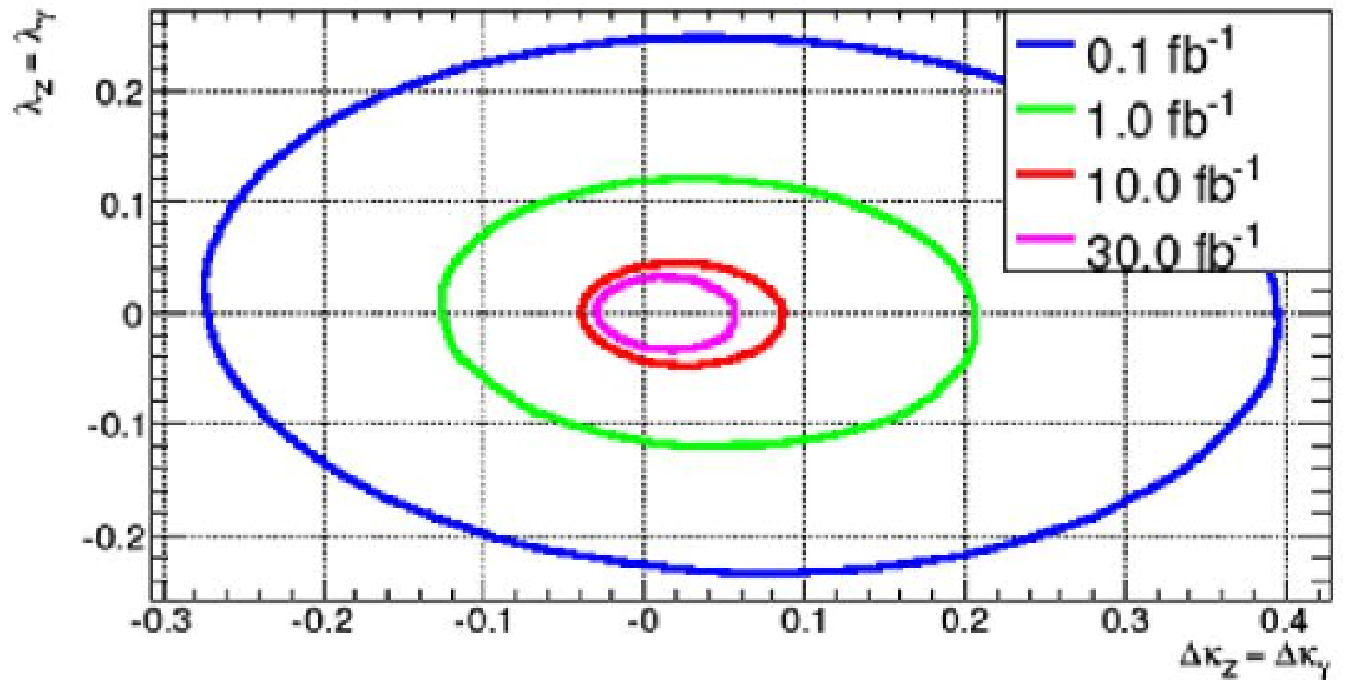,height=1.7in}
\caption{Limit on anomalous coupling at ATLAS for various integrated 
luminosities.}
\label{fig:atlasww}
\end{minipage}
\end{figure}

A special role in electroweak processes, is played by diboson production 
and decay. Necessary in the theory to guarantee unitarity, they represent 
a window on the unknown as $WW$ and $WZ$ are produced through diagrams 
involving trilinear gauge couplings.
However, at the Tevatron, their tiny 
cross sections set them at the boundary of the observable ($\sigma 
(p\bar{p}\rightarrow WZ)\simeq 4$ pb at 1.96 TeV). In 
figure~\ref{fig:dibt} 
you can see the most recent Tevatron results for $WW$ production. Recently 
CDF measured $\sigma (ZZ)=1.56^{+.80}_{-.63}\pm 0.25$ pb in 4.8 fb$^{-1}$ 
and $\sigma(WW)=14.4 \pm 3.1 \pm 2.2$ pb in 3.9 fb$^{-1}$.

The large statistics available at the LHC will allow to set strong limits, 
after a few years, on new physics. As an example 
figure~\ref{fig:atlasww} shows the ATLAS 
expectations for anomalous couplings in the $WW$ channel, looking for
anomalous couplings in the $WWZ$ and $WW\gamma$ vertices. 

\subsection{$W$ mass}
This (free) parameter of the SM has a specific relevance as, combined with 
the mass of the top quark, it is related, through loop diagrams, to the 
mass of the Higgs particle. 

\begin{figure}
\begin{minipage}{0.45\linewidth}
\psfig{figure=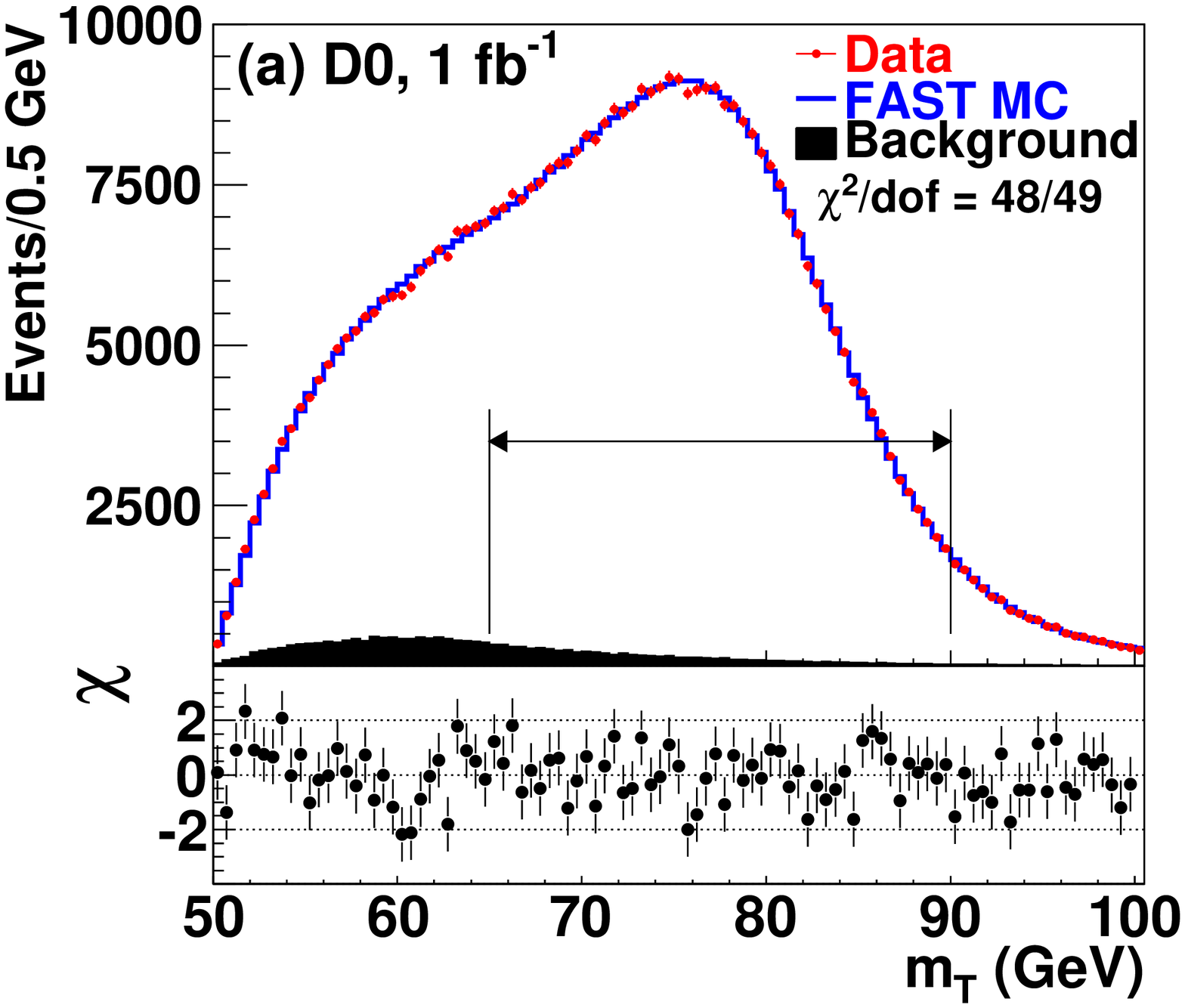,height=1.7in}
\caption{D0 transverse mass for data and MC. Fit residuals also shown in 
inset.}
\label{fig:d0wm}
\end{minipage}
\begin{minipage}{0.45\linewidth}
\psfig{figure=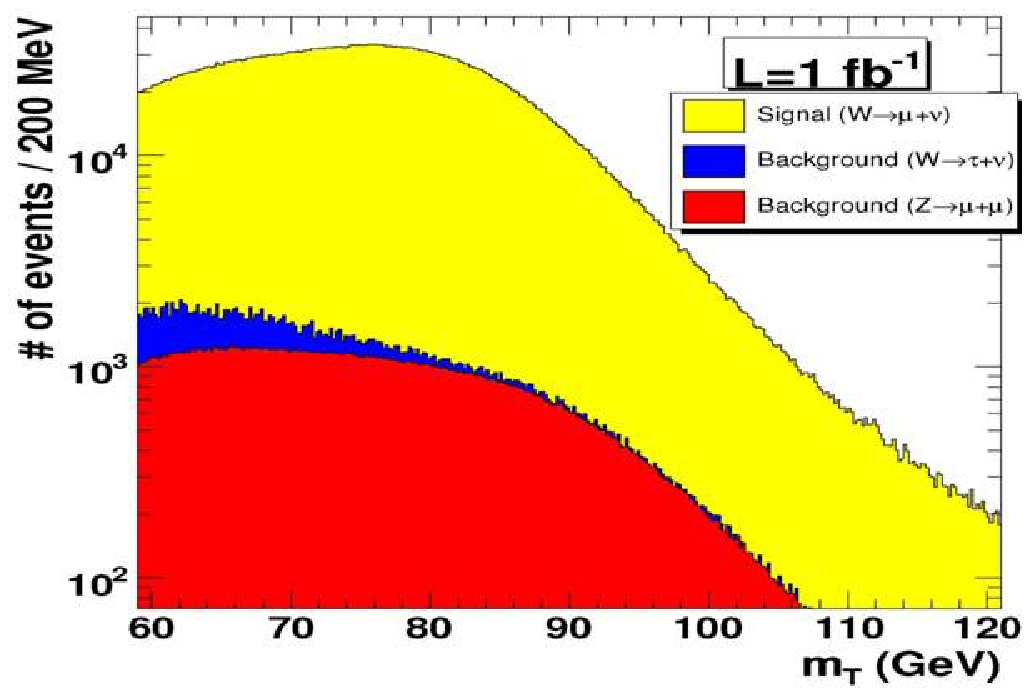,height=1.7in}
\caption{CMS signal for $M_W$ in 1 fb$^{-1}$, 14 TeV.}
\label{fig:mwcms}
\end{minipage}
\end{figure}

The $W$ mass in measured in events where the boson decays into $\nu$ and $e$ or $\mu$.
CDF, in 0.2 fb$^{-1}$ measures $80.431\pm 0.034(stat) \pm 0.034 (syst)$ GeV/c$^2$, 
D0 in 1 fb$^{-1}$ 
$80.401\pm 0.021(stat)\pm 0.038(syst)$ GeV/c$^2$ (see 
figure~\ref{fig:d0wm}).
Their combined result (80.420 GeV/c$^2$) has an uncertainty of only 31 
MeV/c$^2$,
an improvement over LEP uncertainty (33 MeV/c$^2$) where, however, four 
experiments were involved. This brings the World Average to $80.399\pm 
0.023$ GeV/c$^2$. CDF is working to update its result with
$\sim$ 2 fb$^{-1}$. 
The expected statistical accuracy is $\simeq 16$ MeV/c$^2$ per each 
channel ($e$, $\mu$).

Both ATLAS and CMS plan on determining $W$ mass with an accuracy of 
$\simeq$ 10 MeV/c$^2$.
At the moment, it looks like this measurement is a long way to go for the LHC.
A tough systematics, to be throughly undrestood, forces to reject 
most of the events. As an example the recent D0 result uses only $\approx$ 
20 \% of the $W$'s produced. 
A recent estimate by ATLAS (15 pb$^{-1}$ at 14 TeV) predicts 
uncertainties, statistical and systematics combined, O(200) 
MeV/c$^2$, still far from the Tevatron results while CDF and D0 project a 
$\approx$ 10 MeV/c$^2$ limit for a run with 10 fb$^{-1}$. A prediction 
that, however, seems very ambitious at this time.

\section{Top Physics}

Since its discovery in 1995, top is a real focus for the Tevatron physics 
programme as Fermilab is the only place where it can be studied.
Thanks to its large mass, top decays before hadronization, therefore 
provides the 
two experiments with a unique place to test QCD prediction. 
Its peculiarity 
easies the comparison of measured production cross section with 
predictions and makes the top quark mass one of the most accurate 
measurement of the SM parameters. 

At LHC  $t\bar{t}$ pairs are produced with a cross section of 
500(800) pb at 10(14) TeV. Therefore tens of thousands of events will be 
available soon after the start of operation. While many studies focus on 
the use of those events to improve understanding of the detectors, there 
are physics measurements that will exploit at best this "top 
factory".
Indeed we will see that the large event yield opens
interesting perspectives to
shed light on the 3$^{rd}$ family couplings.

\subsection{Production and decays}

Until recent times, top quark has been observed only in strong 
production of $t\bar{t}$ pairs. This process proceeds for about 85(15) \% 
through quark (gluon) fusion at the Tevatron. The situation, thanks to the 
larger energy available at parton level, reverses at the LHC.

As $t$ decays $\simeq$ 100 \% in $W$ and a $b$ quark, different channels 
for $t\bar{t}$ are classified (and named) after the $W$ decays. 
The most important ones, thanks to a combination of branching fractions, 
and of
signal over backgroud ratio, are the dilepton channel (where both $W$'s 
decay into leptons) and the lepton$+$jets channel, where one of the $W$ 
decays into two jets. To improve $S/B$ in the latter case, $b$-tagging 
(i.e. the positive identification of at least one jet as coming from the 
hadronization of a $b$ quark) is applied. Charged leptons mostly used 
for those measurements are $e$ and $\mu$, with $\tau$ playing a lesser 
role.
\begin{figure}
\begin{minipage}{0.45\linewidth}
\psfig{figure=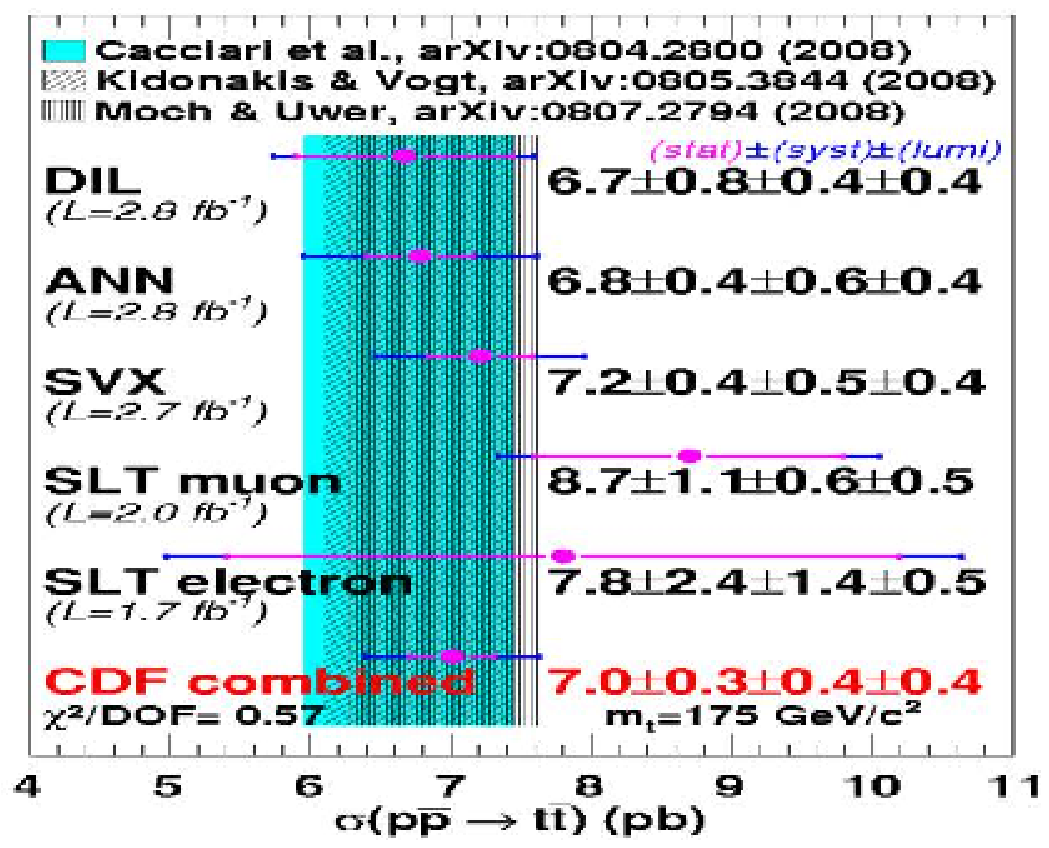,height=1.7in}
\caption{$\sigma_{t\bar{t}}$ measured by CDF compared to NLO 
predictions.}
\label{fig:tnlo}
\end{minipage}
\begin{minipage}{0.45\linewidth}
\psfig{figure=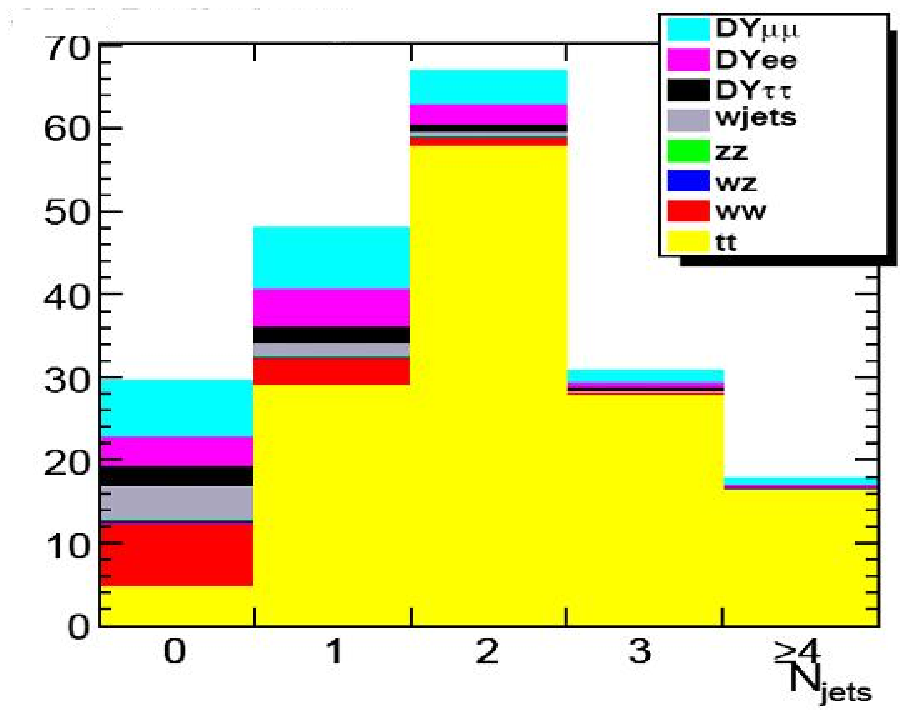,height=1.7in}
\caption{$W+$ jets distribution in $t\bar{t}$ events at CMS.}
\label{fig:cmstt}
\end{minipage}
\end{figure}

CDF recently updated its results in the dilepton and lepton+jets (and $b$ 
tagging) channels with $\approx 5$ 
fb$^{-1}$. The measured cross section (assuming $M_{top}=172.5$GeV/c$^2$) 
is $7.5\pm 0.31(stat)\pm 0.34(syst)\pm 0.15(Z\ th.+ residual\ lum.)$ pb. 
The Tevatron results already challenge the theoretical predictions as you 
can see in figure~\ref{fig:tnlo} where most recent NLO calculations 
(uncertainty O(10\%)) are compared to the CDF experimental determination.

Both ATLAS and CMS studied different channels in various scenarios.
At CMS, for the dilepton channel, a 10\% precision on $\sigma_{t\bar{t}}$ 
is predicted with as little as 10 pb$^{-1}$ at 14 TeV.
Figure~\ref{fig:cmstt} shows the expected $W+$jet distribution for this 
channel ($ee$,$e\mu$, $\mu \mu$).
ATLAS predicts an accuracy $\Delta \sigma/\sigma \simeq 4.5(stat) 
\oplus 8(syst)$\% using $b$-tagging with 100 pb$^{-1}$ of data at 14 TeV.
These preliminary figures are very encouraging. Besides, while channels 
with $\tau$ were of little use at the Tevatron, the large statistics might 
allow them to play a more significant role at the LHC. A possibility to be 
explored is to determine $M_{top}$ from a precise cross section 
determination as there is a mild dependence of $\sigma$ upon the top 
mass.

\begin{figure}
\begin{minipage}{0.45\linewidth}
\psfig{figure=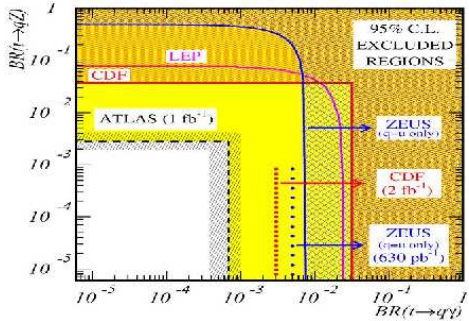,height=1.7in}
\caption{ATLAS expectations in 1 fb$^{-1}$ for FCNC top decays.}
\label{fig:atlasfcnc}
\end{minipage}
\begin{minipage}{0.45\linewidth}
\psfig{figure=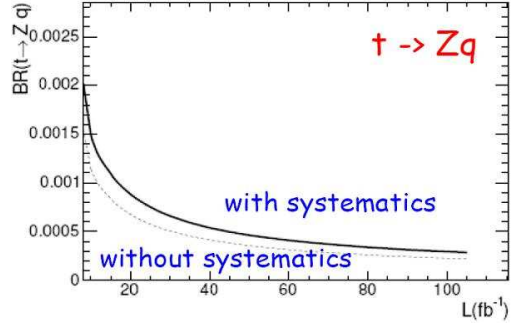,height=1.7in}
\caption{CMS expectation for $Zq$ limits as a function of $\int Ldt$.}
\label{fig:cmszq}
\end{minipage}
\end{figure}

The large statistics can be very helpful also in exploring aspects of top 
physics linked to the decay channels. CDF pioneered the search for FCNC 
currents ($t\rightarrow Zq$ and $t\rightarrow q\gamma$) setting limits 
$\simeq 3\div 4$\%. While these channels are, of course, suppressed at 
tree level in SM, in some theories they appear at the O(10$^{-4}$) level.
In figure~\ref{fig:atlasfcnc} the result of ATLAS study using 1 fb$^{-1}$ 
of data compared to current limits. FCNC decays were also studied by CMS as a 
function of the collected integrated luminosity (figure~\ref{fig:cmszq}). 
With the large statistics available, these decays represent a real window 
on scenarios beyond the SM.


\subsection{Top Mass}

As mentioned earlier, $W$ mass is linked, through (logarithmic) loop corrections 
involving the 
mass of the top quark, to the Higgs sector. Therefore, within the SM, 
these two measurements provide an hint on the Higgs mass and can suggest 
the emergence of new physics (if direct and indirect measurements show 
some strain).

The measurements of the top mass typically proceed through the comparison 
of mass-sensitive observables in data and MC.
At the end of Run I, D0 pioneered a technique which, using as 
additional information the dynamics of $t\bar{t}$ events, 
improved the statistical accuracy~\cite{d0nat}. Nowadays this path  is 
largely exploited by both experiments.

\begin{figure}
\begin{minipage}{0.45\linewidth}
\psfig{figure=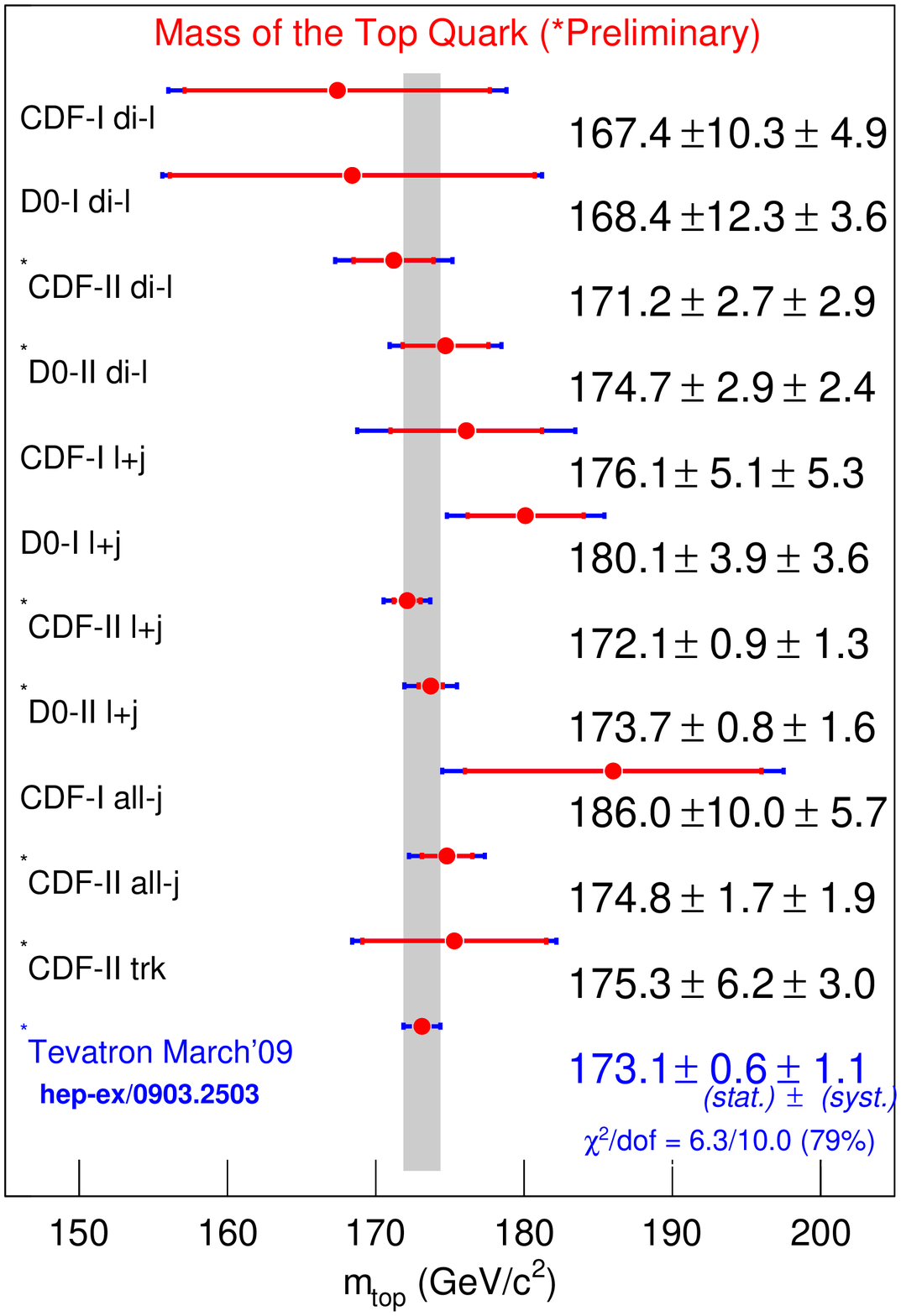,height=1.7in}
\caption{Top Quark mass as measured at the Tevatron.}
\label{fig:tmcdf}
\end{minipage}
\begin{minipage}{0.45\linewidth}
\psfig{figure=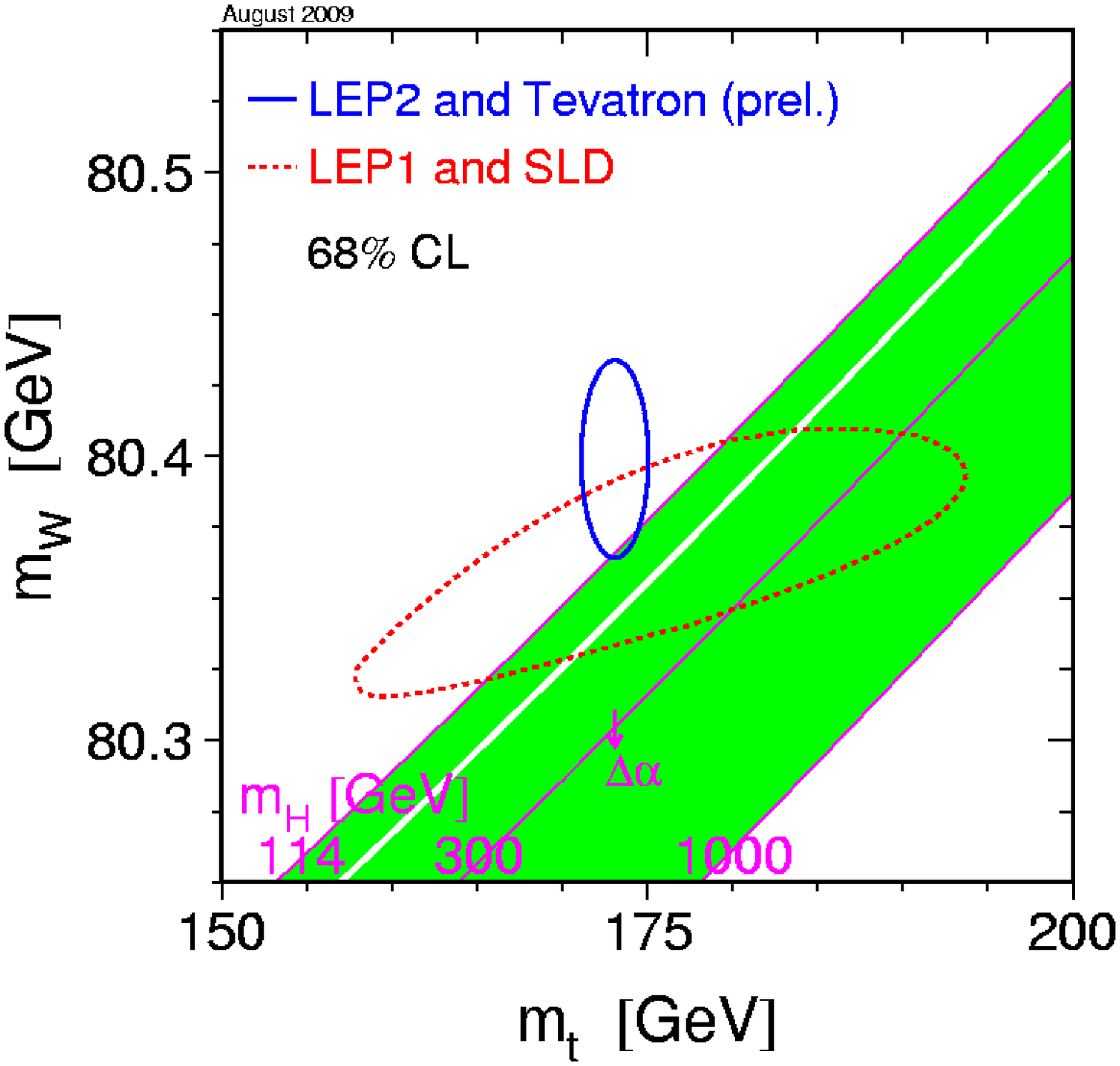,height=1.7in}
\caption{$M_W$-$M_{top}$ plane: data compared to expectation for different 
$M_H$.}
\label{fig:tmhiggs}
\end{minipage}
\end{figure}

At the Tevatron both CDF and D0 measure $M_{top}$ in various channels:
dilepton, l+jets and all-hadronic. Only the latter has a 
fully-reconstructed final state, the others contain at least one 
neutrino 
with an unknown $P_z$. In all cases, as top and $W$ decays in quarks, 
hadronization corrections are needed to go back from the (measured) jet 
energies to the initial parton energies.Together with other energy-related
uncertainties, they are dubbed Jet Energy Scale (JES) and 
represent the most important systematics to measure $M_{top}$.
In order to obtain an {\em in situ} calibration of the JES, $W\rightarrow jj$ in 
$t\bar{t}$ events is used as an additional constraint. This reduces this systematic
error as a function of increased luminosity~\cite{topm}.
In figure~\ref{fig:tmcdf} I present a full compilation of the results 
obtained in various channels. The 
current Tevatron average, obtained with 3.6 fb$^{-1}$ of data, is $173.1 
\pm 0.6 (stat) \pm 1.1 (syst)$ 
GeV/c$^2$. A striking determination with 
an accuracy of better than 1 \% which makes $M_{top}$
the best known of 
all quark masses. However, at this level, there is a debate among 
theorists on how to interpret this accuracy.
Nevertheless, when put together with the mass of the $W$ it sets a strong 
constraint on the Higgs sector (see figure~\ref{fig:tmhiggs}). As of 
today, the 95 \% 
C.L. limit obtained by indirect EWK fits is: $M_H<157$ 
GeV/c$^2$~\cite{higgsd}.

\begin{figure}
\begin{minipage}{0.45\linewidth}
\psfig{figure=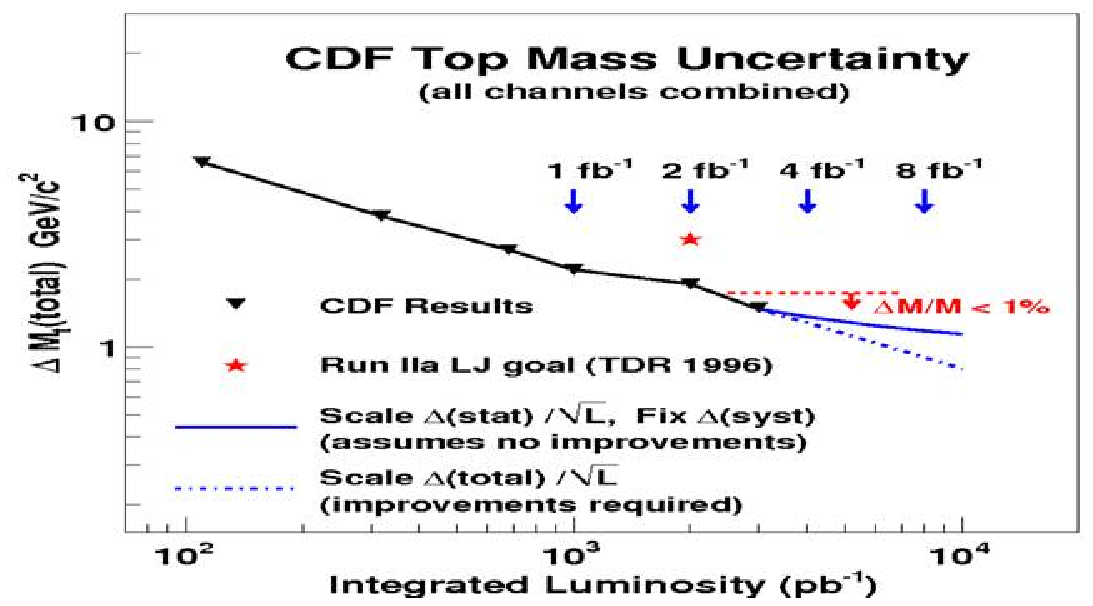,height=1.5in}
\caption{CDF projections on $M_{top}$.}
\label{fig:tmfut}
\end{minipage}
\begin{minipage}{0.45\linewidth}
\psfig{figure=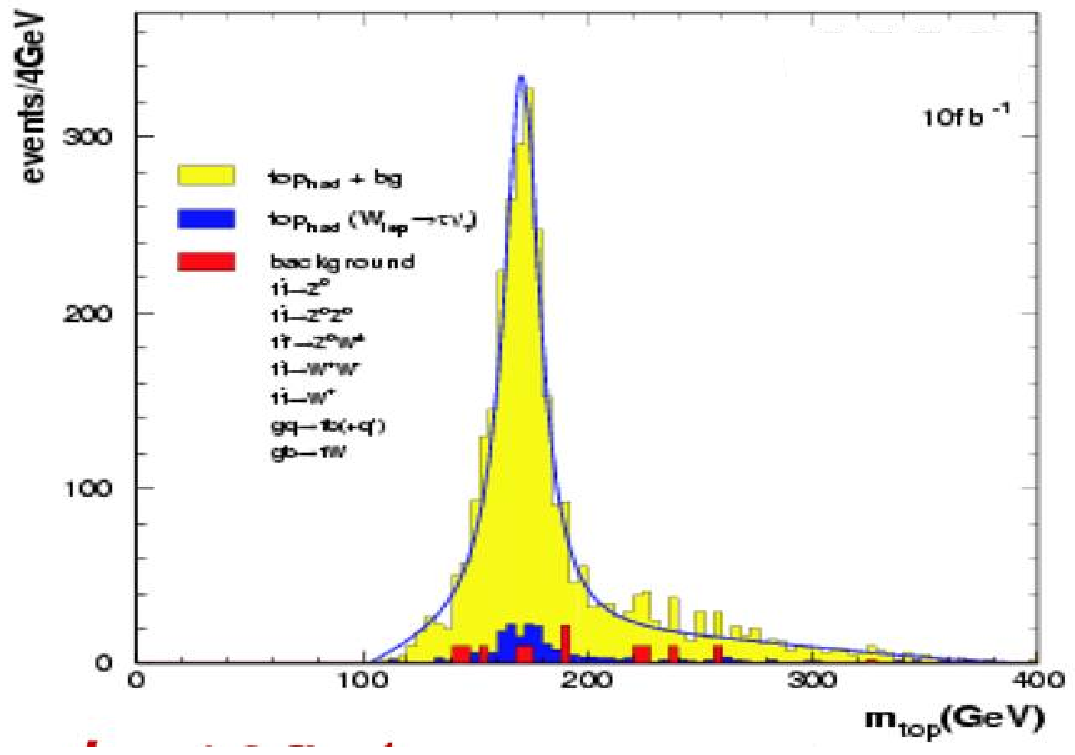,height=1.5in}
\caption{$M_{top}$ at CMS.}
\label{fig:cmstm}
\end{minipage}
\end{figure}

The future looks still bright for the Tevatron as shown in 
figure~\ref{fig:tmfut}, where
you can see CDF 
expectations as a function of the collected integrated luminosity. 
Even without improvements in the analysis, luminosity will 
provide a chance to possibly reach the 1 GeV/c$^2$ limit.

The LHC experiment will exploit techniques and studies 
performed at the Tevatron.
As the current predictions by CMS (see figure~\ref{fig:cmstm}) and ATLAS 
are
$\delta M_{top}\approx 1 GeV/c^2$ with 10 fb$^{-1}$ at 14 TeV,
I expect that the top mass will be a lasting heritage of the 
Tevatron. 

\subsection{Single Top}
One of the most interesting electroweak processes, is the production of 
events containing one top quark. There are three relevant Feynman 
diagrams: $s$-channel, $t$-channel, associated production. In the 
$s$-channel one $W$ is produced by two light quakrs 
and decays in a $t$ and a $b$ quark. In the $t$ channel the top quark is 
produced in association with a light quark and finally the the 
associated production produce a final state containing a $t$ quark and a 
$W$. At the Tevatron the latter has a negligible ($\approx 0.3$ pb) cross 
section, while the $s$ channel is $\sigma = 0.88 \pm 0.11$ pb and the $t$ 
channel $1.98 \pm 0.25$ pb at NLO. One of the reason why this 
channel is interesting is that, due to its peculiar production, 
$\sigma_{single-t} 
\propto |V_{tb}|^2$, therefore a direct measurement 
of this CKM matrix element can be performed.

\begin{figure}
\begin{minipage}{0.45\linewidth}
\psfig{figure=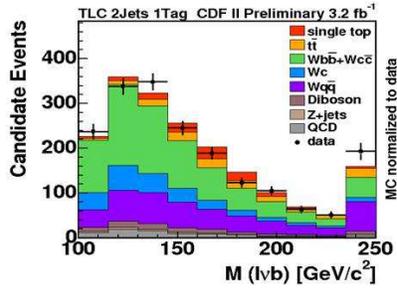,height=1.5in}
\caption{$M_{l\nu b}$ for all events with 2 jets, $\geq 1$ b-tag.}
\label{fig:stall}
\end{minipage}
\begin{minipage}{0.45\linewidth}
\psfig{figure=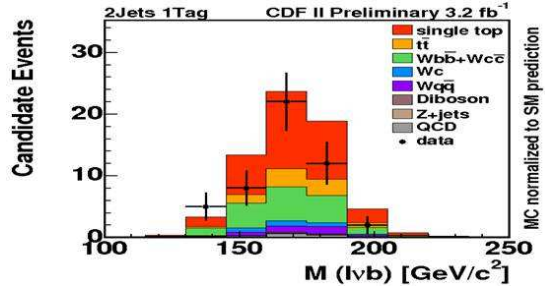,height=1.5in}
\caption{$M_{l\nu b}$ for selected events, $\geq 1$ b-tag.}
\label{fig:stsel}
\end{minipage}
\end{figure}

The tiny cross section is a formidable problem at the Tevatron. Despite 
the large integrated luminosity in Run II, it still took 14 years from  
$t\bar{t}$ to $s$-top discovery. Indeed the final state of single 
top events is represented by events containing a $W$ and $2$ jets 
(contatining one or two $b$ jets). 
We know that the 
large $W+b\bar{b}$ and $Wjj$ generic events constitutes a significant 
background. Moreover the standard $t\bar{t}$ production constitues another 
source of background.
Overall, even after all selection requirements, the signal is a fraction 
of background (S/B $\sim 1/16$) and a counting experiment is not possible.
CDF and D0 used a number of
multivariate separation techniques that, exploiting the excellent 
knowledge of the detectors response and of the backgrounds, allow to 
statistically separate signal-like events from the overwhelming 
background.
As an example I show in figure~\ref{fig:stall} and~\ref{fig:stsel} the 
distribution of 
the invariant mass of the system $l\nu b$ for all events and for events 
selected as signal-like by a Neural Network. It is clear that, in the 
second plot, data favours the presence of events containing single top.
CDF published $\sigma = 2.3^{+0.6}_{-.05}$ pb and D0 $\sigma =3.94 \pm 
0.88$ pb. The combined Tevatron result is $2.76^{+0.58}_{-0.41}$ pb.
From these measurements one can solve for the CKM 
element $V_{tb}= 0.88 \pm 0.07$, or $|V_{tb}|>0.77$ at 95 \% C.L..

At the LHC single top is produced through the same mechanisms but the 
cross sections are quite different:$s$-channel: 11 pb, $t$-channel: 250 
pb, associated production: 66 pb.
$t$-channel appears to be the best candidate for observation, however the 
large background due to $W+$jets events requires $b$-tagging to improve 
S/B. 
A recent study by ATLAS found the original TDR to be optimistic and, with 
a cut based selection they now estimate S/B$\simeq 1/3$ in 1fb$^{-1}$.
In the same amount of data, using "multivariate methods" as D0 and CDF, it 
improves to be $\simeq 1.3$ with an expected accuracy $\Delta 
\sigma/\sigma$ 
$\sim 22$\%, more or less the same as obtained at the Tevatron.

\section{Conclusion}
Precision electroweak measurements and top physics are the basics of 
physics at hadron colliders.
However, despite they belong to the realm of standard model physics, many 
interesting topics can be covered and new physics can be explored.
Their full understanding is the key to proceed forwards in the realm of 
the unknown. Tevatron experiments showed that unexpected accuracies can be 
obtained thanks to the ingenuity of the physicists involved. Measurements 
of $M_{top}$ and $M_W$ are good examples. 

As there are still many results which are statistically limited, we 
expect significant contribution by ATLAS and 
CMS "soon". They will operate at the border of the standard model where we 
are now testing fundamental aspects of the theory. 

Be ready for suprises as we move to new energies.

\section*{Acknowledgments}

I would like to acknowledge my D0 and CDF colleagues for providing with 
plenty of information on the ongoing analyses. A special thanks to
V.~Vercesi and R.~Tenchini for providing helpful information on the
ongoing analyses at ATLAS and CMS. All mistakes are, of course, mine.

\end{document}